\documentclass{moriond}

\def\Journal#1#2#3#4{{#1} {\bf #2}, #3 (#4)}

\def\PRL{\em Phys. Rev. Lett.}
\def\PRD{{\em Phys. Rev.} D}

\def\NAT{{\em Nature}}

\def\be{\begin{equation}}
\def\ee{\end{equation}}
\def\bea{\begin{eqnarray}}
\def\eea{\end{eqnarray}}

\newcommand{\lm}{{\ell m}}
\usepackage{multirow}
\usepackage[T1]{fontenc}
\usepackage[utf8]{inputenc}
\usepackage[english]{babel}
\usepackage[svgnames,dvipsnames,table]{xcolor}

\usepackage{graphicx,float}
\usepackage{booktabs} 
\usepackage[font=small,labelfont=bf]{caption} 
\usepackage{amsfonts, amsmath, amsthm, amssymb, units} 
\usepackage{wrapfig} 
\usepackage{hyperref,url}
\usepackage{times}
\usepackage{tcolorbox}
\usepackage{dashrule}
\usepackage{multicol} 
\usepackage{bbold}
\usepackage{tcolorbox}
\usepackage{subfig}
\usepackage{enumitem}
\usepackage{algorithm}
\usepackage{makecell}
\usepackage{array}
\usepackage{epsfig}
\usepackage{graphics}
\usepackage{graphicx}
\usepackage{bm}
\usepackage{amssymb}
\usepackage{xspace}
\usepackage[normalem]{ulem} 
\usepackage{url}
\usepackage{float}
\usepackage[bottom]{footmisc}
\usepackage{lineno}
\usepackage{mathrsfs}
\usepackage{makecell}
\usepackage{microtype}
\usepackage{mathtools}

\newcommand{\Photo}{}

\begin{document}
\title{Tests of general relativity in the nonlinear regime:\\
a parametrized plunge-merger-ringdown waveform model}

\author{E. Maggio}

\address{Max Planck Institute for Gravitational Physics (Albert Einstein Institute), D-14476 Potsdam, Germany}

\maketitle

\abstracts{
Gravitational waves (GWs) provide a unique opportunity to test gravity in the dynamical and nonlinear regime. We present a parametrized test of general relativity (GR) that introduces generic deviations to the plunge, merger and ringdown stages of binary--black-hole coalescences. The novel feature of the model is that it can capture signatures of beyond-GR physics in the plunge-merger phase. We use the model to provide constraints on the plunge-merger parameters from the analysis of GW150914. Alarmingly, we find that GW200129 shows a strong violation of GR. We interpret this result as a false violation of GR either due to waveform systematics (mismodeling of spin precession) or data-quality issues.}

\section{Parametrized plunge-merger-ringdown model}

\begin{figure}[t]
\begin{center}
\includegraphics[scale=0.5]{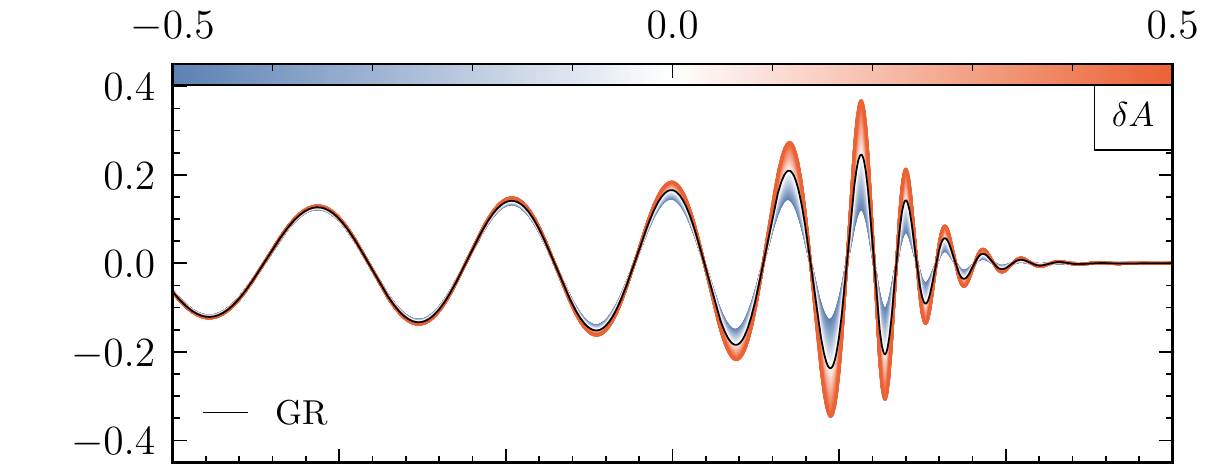}\\
\includegraphics[scale=0.5]{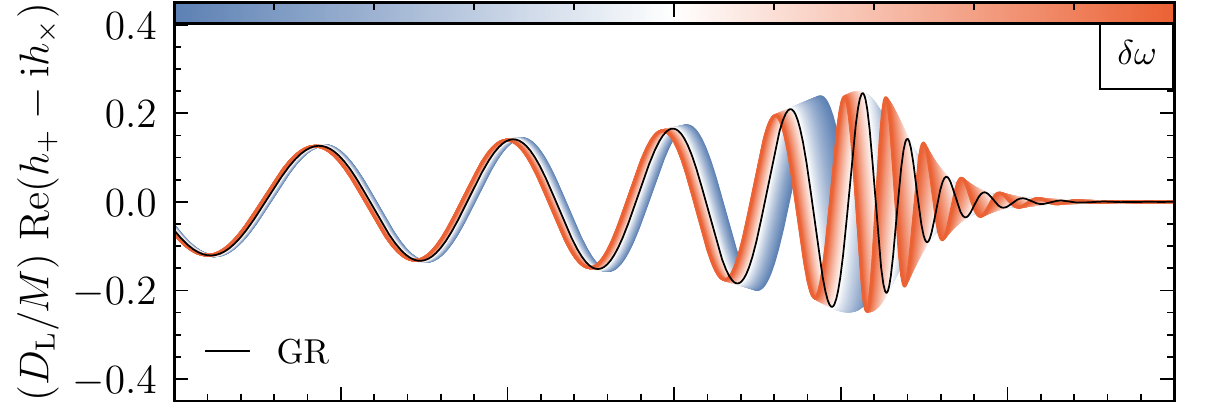}\\
\includegraphics[scale=0.5]{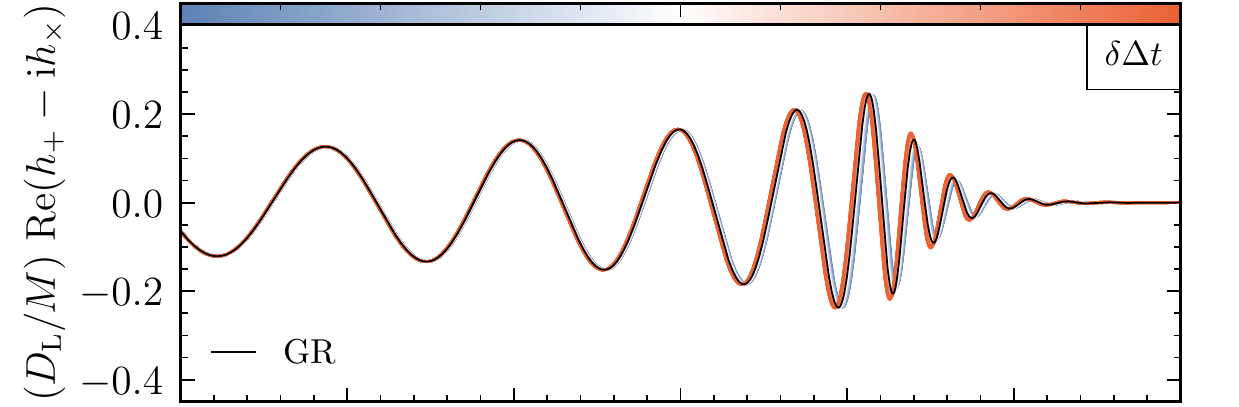}\\
\includegraphics[scale=0.5]{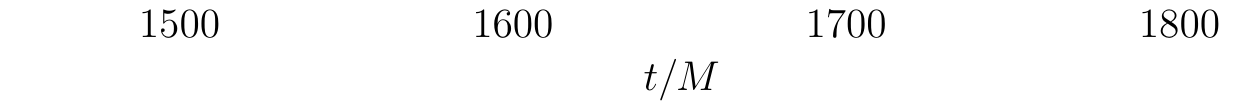}
\end{center}
\caption{Time evolution of the GW strain for non-zero values of the plunge-merger parameters.}
\label{fig:params}
\end{figure}

We use as our baseline model the waveform constructed in the effective-one-body formalism for quasicircular and spin-aligned binaries, which is calibrated to numerical relativity simulations and contains high-order modes,
i.e., \texttt{SEOBNRv4HM$\textunderscore$PA}~\cite{Buonanno:1998gg,Buonanno:2000ef,Bohe:2016gbl,Cotesta:2018fcv,Nagar:2018gnk,Mihaylov:2021bpf}.
The waveform is constructed by attaching the merger-ringdown waveform to the inspiral-plunge waveform at a matching time,
\begin{equation}
    h_{\ell m}(t) = h_{\ell m}^{\rm insp-plunge}(t) \, \Theta \left( t_{\rm match}^{\ell m}-t \right)
    + h_{\ell m}^{\rm merger-RD}(t) \, \Theta \left( t-t_{\rm match}^{\ell m} \right) \,,
\end{equation}
where $\Theta(t)$ is the Heaviside function, and the matching time is defined as
\begin{equation}
t_{\rm match}^{\ell m} =
\begin{cases}
	 t^{22}_{\rm peak} \,, &(\ell,m)=(2,2), \, (3,3),\, (2,1), (4,4) \,, \\
	 t^{22}_{\rm peak} - 10M \,, &(\ell,m)=(5,5) \,,
\end{cases}
\end{equation}
where $t^{22}_{\rm peak}$ is the time in which the amplitude of the $(2,2)$ mode peaks. The amplitude and the orbital frequency peak at different times, therefore we introduce a time-lag parameter $\Delta t^{\rm{GR}}_{\ell m} = t^{\Omega}_{\rm peak}- t^{22}_{\rm peak}$.
We develop a parametrized waveform model that allows for deviations from GR in the plunge-merger-ringdown stage~\cite{Maggio:2022hre}. We introduce fractional deviations to the NR-informed amplitude and frequency at the matching time and the time-lag parameter as,
\begin{eqnarray}
    |h_{\lm}^{\rm NR}| &\to& |h_{\lm}^{\rm NR}| \, (1 + \delta A_{\lm})\,,
    \\
    \omega_{\lm}^{\rm NR} &\to& \omega_{\lm}^{\rm NR} \, (1 + \delta \omega_{\lm})\,, \\
    \Delta t^{\rm GR}_{\ell m} &\to& \Delta t^{\rm GR}_{\ell m} \, \left( 1 + \delta \Delta t_{\ell m} \right) \,.
\end{eqnarray}
We also introduce deformations to the quasi-normal modes following the same strategy applied in Refs.~\cite{Brito:2018rfr,Ghosh:2021mrv} for the oscillation frequency and damping time of the fundamental modes,
\begin{eqnarray*}
    f_{\lm 0} &\to f_{\lm 0} \, (1 + \delta f_{\lm 0}) \,,
    \\
    \tau_{\lm 0} &\to \tau_{\lm 0} \, (1 + \delta \tau_{\lm 0}) \,.
\end{eqnarray*}
For simplicity, we assume that the plunge-merger parameters have the same values across different modes (i.e., $\delta A_{\lm} = \delta A$, $\delta \omega_{\ell m} = \delta \omega$, $\delta \Delta t_{\ell m} = \delta \Delta t$), and the ringdown parameters are nonzero only for the fundamental $(2,2)$ mode.
Figure~\ref{fig:params} shows the waveform morphology as a function of  time where the plunge-merger parameters vary in the range $[-0.5,0.5]$ for 
$\delta A$ (top panel), $\delta \omega$ (middle panel) and $\delta \Delta t$ (bottom panel). The GR waveform is shown by the black solid line.
%

\section{Bayesian parameter estimation}

We analyse the events GW150914~\cite{LIGOScientific:2016aoc} and GW200129~\cite{LIGOScientific:2021djp}, which are among the loudest binary--black-hole signals to date, by varying the binary parameters and a subset of the deviation parameters, i.e., $\{ \delta A, \delta \omega \}$, to perform a ``plunge-merger test of GR''.
From the analysis of GW150914, we provide constraints on the deviations from GR, i.e., $\delta A = -0.01^{+0.27}_{-0.19}$ and $\delta \omega = 0.00^{+0.17}_{-0.12}$~\cite{Maggio:2022hre}. Deviations in the intermediate region of the \texttt{IMRPhenom} waveform were  constrained in the TIGER test~\cite{LIGOScientific:2016lio}.
From the analysis of GW200129, the inferred value of $\delta \omega$ $\left(\delta \omega =
-0.002^{+0.097}_{-0.082}\right)$ is consistent with GR, however the inferred
value of $\delta A$ $\left(\delta A = 0.44^{+0.38}_{-0.28}\right)$ exhibits a gross violation of GR.
The apparent deviation from GR could be due to systematic errors in the GW modeling or data-quality issues as highlighted by the analyses of Refs.~\cite{Hannam:2021pit,Payne:2022spz}. 
We explore the former possibility by performing synthetic-data studies where the injected GW signal is generated with spin-precessing waveform models. 
We conclude that the presence of spin-precession could bias us to find a false evidence for beyond-GR effects when using a nonprecessing non-GR model.
%

\section*{Acknowledgments}

%
We thank Alessandra Buonanno and Hector O. Silva for discussions, and Gregorio Carullo for comments.
%
%
We acknowledge funding from the Deutsche Forschungsgemeinschaft (DFG)~-~project number:~386119226.
%
%
%
%
The material presented in this paper is based upon work supported by National
Science Foundation's (NSF) LIGO Laboratory, which is a major facility fully
funded by the NSF.

\section*{References}

\end{document}